\documentclass{PoS}

\title{Unifying disc-jet behaviour in X-ray binaries: an optical/IR approach}

\ShortTitle{Optical/IR disc-jet behaviour in XBs}

\author{\speaker{David M. Russell}\\
        University of Amsterdam\\
        E-mail: \email{D.M.Russell@uva.nl}}

\author{Dipankar Maitra\\
        University of Amsterdam\\
        E-mail: \email{D.Maitra@uva.nl}}

\author{Rob P. Fender\\
        University of Southampton\\
        E-mail: \email{r.fender@soton.ac.uk}}

\author{Fraser Lewis\\
        Las Cumbres Observatory Global Telescope Network / Open University\\
        E-mail: \email{lewisf@Cardiff.ac.uk}}

\abstract{
Synchrotron emission from jets produced by X-ray binaries can be detected at optical and infrared (IR) frequencies.  I show that optical/IR colour-magnitude diagrams of the outbursts of nine X-ray binaries successfully separate thermal disc emission from non-thermal jet emission, in both black hole and neutron star sources.  A heated single-temperature blackbody is able to reproduce the observed relations between colour and magnitude, except when excursions are made to a redder colour than expected, which is due to jet emission.  The general picture that is developed is then incorporated into the unified picture of disc-jet behaviour in black hole X-ray binaries.  At a given position of a source in the X-ray hardness-intensity diagram, the radio, IR and optical properties can be inferred.  Similarly, it is possible to predict the X-ray and radio luminosities and spectral states from optical/IR monitoring.
}

\FullConference{VII Microquasar Workshop: Microquasars and Beyond\\
		 September 1-5 2008\\
		 Foca, Izmir, Turkey}

\begin{document}

\section{Introduction}

X-ray emission from low-mass X-ray binaries (LMXBs) typically originates in the hottest regions in the inner accretion flow -- the inner accretion disc and the hot Comptonised region. Radio emission from X-ray binaries manifests itself as synchrotron emission from relativistic jets of outflowing matter and energy. In the last few years it has been realised that optical and infrared (IR) emission from these systems can originate not only from the companion star and the cool, outer regions of the accretion disc, but also from the inner region of the jets (in both black hole and neutron star XBs) \cite{corbfe02,gileet05}. Studies of the variability and spectral shape of the spectrum of the jet in the optical/IR (OIR) have shown it to be optically thin synchrotron emission, with spectral indices typically of values around $\alpha \sim -0.7 \pm 0.1$ where $F_{\nu} \propto \nu^{\alpha}$ \cite{miglet06,hyneet06}. In comparison, the accretion disc produces a thermal spectrum that is blue at OIR wavelengths, with empirical spectral indices around $\alpha \sim + 1.0 \pm 0.5$ \cite{hyne05} during outburst.

Studies of the outer accretion disc can constrain its geometry, the effect of X-ray irradiation on it and other properties \cite{cunn76,franet02}. By studying optically thin synchrotron emission from the jet, constraints can be made of the total power of the jet and the local conditions close to where the jet is launched \cite{market01,russfe08}. For both the disc and the jet, a wealth of information can be derived from how the emission properties evolve with changes in mass accretion rate, i.e. changes in luminosity during an outburst. Since the main two OIR contributors during LMXB outbursts are the disc and the jet \cite{russet06}, which are expected to have different spectral indices, OIR colour changes can disentangle the two emission processes and how their contributions evolve during outburst cycles. It has recently been shown \cite{maitba08} that the relationship between OIR colour and flux during eight outbursts of the neutron star LMXB Aql X--1 is consistent with a single-temperature blackbody heating up, i.e. from the irradiated accretion disc. We expect OIR colour changes when X-ray state transitions are made if the jet contributes, because the jet (both its radio and IR counterparts) is quenched during soft X-ray states, at least in the black hole systems \cite{gallet03,miglfe06}. Rather surprisingly, few analyses of OIR colour changes during LMXB outbursts exist to date, but those that do exist have provided interesting informations about the properties of these emitting regions \cite{chevil95,buxtba04,maitba08}.

The dependence of spectral shape with flux in the X-ray regime has proven pivotal in understanding the coupling between the inflow and outflow in XBs. Here, the X-ray spectral components and timing properties, and radio properties all depend on the position a source lies in the X-ray hardness--intensity diagram (HID), and a hysteretical pattern exists in which the state transition occurs at a higher flux level on the outburst rise than on the decline \cite{maccco03,fendet04}. If OIR spectral and timing properties are also correlated with the position on the X-ray HID, couplings can be inferred between the inner disc (X-ray), the large-scale jet (radio), the outer disc (OIR) and the inner regions of the jet (OIR). If this can be achieved for many XBs, the unification of these systems and their behaviour will be stronger than ever. Introducing time-dependency into models of the behaviour of XBs is vital in understanding them in general (see Maitra et al.; these proceedings for discussions).

\section{Methodology}

We collected optical and IR fluxes of outbursting LMXBs in order to study how the OIR colours evolve with flux. Most of the data are taken from the literature, but some new observations have been added from monitoring campaigns using the Yale 1.0m telescope at Cerro Tololo Inter-American Observatory (CTIO) and the two 2.0m Faulkes Telescopes at Haleakala on Maui (FT North) and Siding Spring in Australia (FT South). The data collection, new observations and data reduction will be detailed in a forthcoming paper (Russell et al. in preparation).

We compare the resulting colour-magnitude diagrams to what we may expect for emission from an accretion disc and for optically thin synchrotron emission from the jet. We use a model for a simple heated blackbody which mimics a single-temperature accretion disc of constant area. This model was successful in reproducing the colour-magnitude diagrams (CMDs) of the outbursts of Aql X--1 \cite{maitba08} (see that paper for details). For the jet, we expect a spectral index of $\alpha \sim -0.7$, which should not change with flux as long as the emission remains optically thin. If the electron distribution in the jet evolves with time we may expect changes in this slope but they will be small in comparison with the colour changes of the irradiated disc. For a disc heated purely by viscous forces (i.e. without irradiation), we expect an OIR spectral index of $\alpha \sim +0.3$ for a multi-temperature blackbody disc, or $+0.3 < \alpha < +2.0$ approaching the Rayleigh-Jeans tail \cite{shaksu73,franet02}. The irradiated disc however can have a large range of values of $\alpha$ including negative values at low flux levels if the blackbody peaks at a longer wavelength than the OIR wavebands \cite{vrtiet90,hyneet02}. The CMDs we construct can therefore separate these different emission processes.

\section{Results}

CMDs of the outbursts of nine LMXBs are shown in Figs. 1, 3 and 4. The only source for which we could obtain optical and IR data of a whole outburst with X-ray state transitions was the black hole XB XTE J1550--564. The CMD of this outburst is presented in Fig. 1 (we use V-band for the optical and H-band for the IR). The heated blackbody model successfully reproduces the trend of the data on the initial rise of the outburst, the soft state, and the final outburst decay. However, at the highest flux levels during the hard state, on both the rise and the decay, there are clear excursions to a redder colour than expected from this model. From the spectral energy distributions it was known that there were two components of OIR emission for this source during the hard state, the redder of which was quenched in the soft state \cite{jainet01,russet06}. This has been interpreted as the jet contribution, which is quenched when the X-ray spectrum softens, like the radio emission. The CMD in Fig. 1 shows that the jet contribution to the OIR dominates at the highest flux levels in the hard state, while the irradiated accretion disc of a constant area likely dominates the OIR at lower fluxes in the hard state, and in the whole of the soft state.

According to correlations between OIR and X-ray emissions for the jet, the irradiated disc and the viscously heated disc in the case of black hole systems \cite{vanpet94,gallet03,russet06}, the jet/disc ratio (the disc here is the irradiated disc) of the OIR emission should increase with increasing flux: $L_{\rm OIR,jet}\propto L_{\rm X}^{0.7}$; $L_{\rm OIR,irrdisc}\propto L_{\rm X}^{0.5}$; $L_{\rm OIR,viscdisc}\propto L_{\rm X}^{0.2}$. Therefore, $L_{\rm OIR,jet}\propto L_{\rm OIR,irrdisc}^{1.4} \propto L_{\rm OIR,viscdisc}^{3.5}$. Assuming these relations are correct, we expect the jet to dominate the OIR emission only at the highest fluxes (depending on its normalisation compared to that of the irradiated disc). Similarly, we would expect the viscous disc to dominate the OIR only at the lower fluxes, since both jet emission and irradiated disc emission rise quicker than this component.

\begin{figure}
\includegraphics[width=15cm,angle=0]{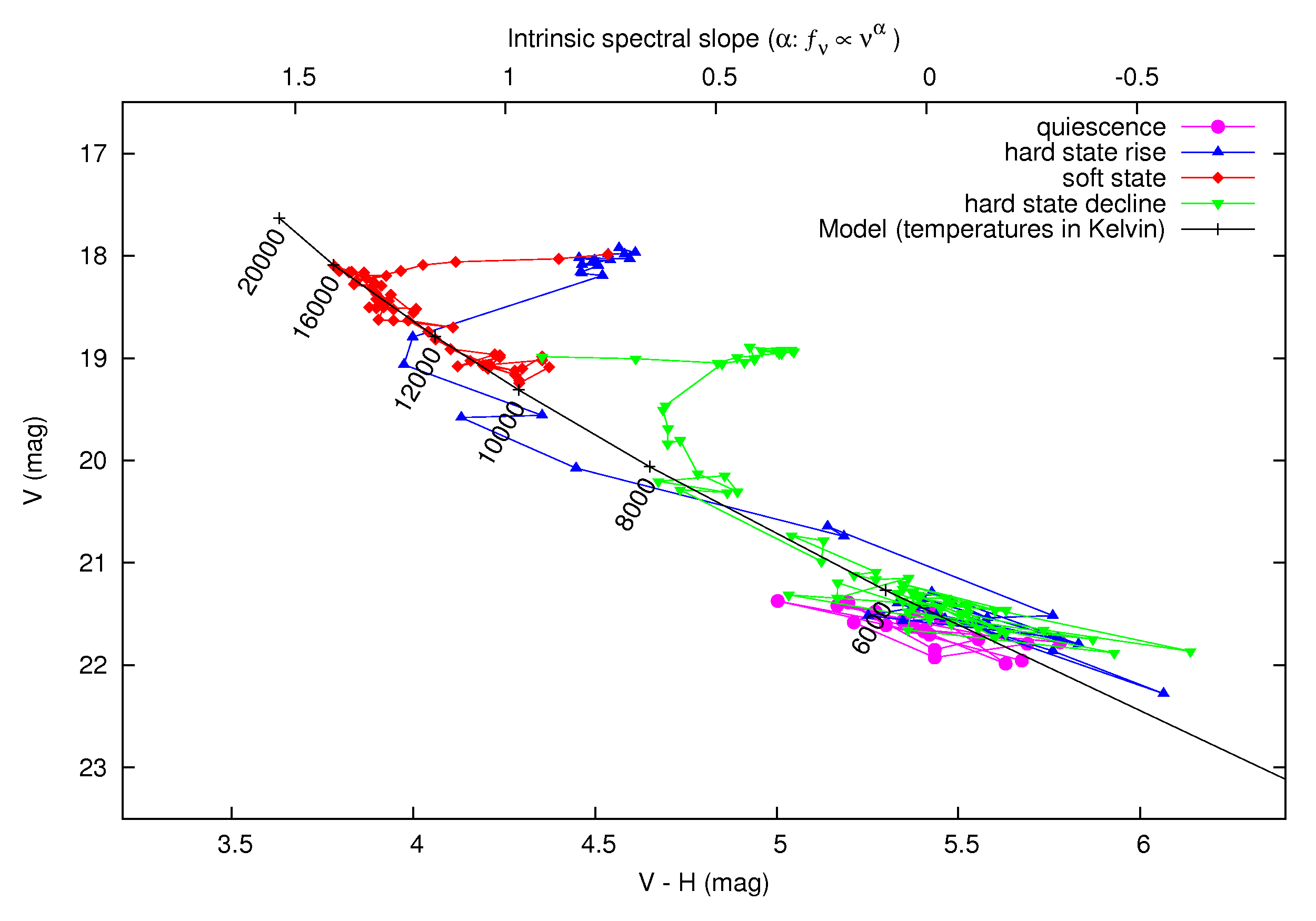}
\caption{OIR CMD for the full outburst of XTE J1550--564 in 2000 (data are from \cite{jainet01}). Overplotted is the model for a heated blackbody. To calculate the intrinsic spectral index we de-redden the magnitudes using estimates of the interstellar extinction $A_{\rm V}$ from the literature.}
\end{figure}

In Fig. 2 we present a schematic representation of what the CMD of an outbursting LMXB may look like given the above assumptions. In panel (a) we show schematics of a CMD given that the OIR emission is dominated by just the irradiated disc (red curve), just the jet (purple curve), or a combination of the two -- one where the jet has a high normalisation (blue) and one where its normalisation is lower (green). For the case of a strong jet (blue curve), the CMD still follows the blackbody relation at low fluxes but rapidly gets redder than this relation at higher fluxes. For the weaker jet (green curve), the OIR spectrum is bluer (following the blackbody relation) before becoming redder at a higher flux than the strong jet case. In panel (c), a jet with a medium normalisation value is shown, with two state transitions. The OIR emission from the jet is instantaneously quenched to zero, while the disc is unchanged. Then the flux is decreased before the jet emission reappears. This plot is able to reproduce most of the CMD of the outburst of XTE J1550--564 (b), except on the outburst decline, where the colour is much redder than the schematic shows. In panel (d), the same schematic is presented but we fix the normalisation of the jet to a higher constant on the decline compared to the rise. This time, the CMD reproduces the hysteresis observed in XTE J1550--564. Therefore, it appears that at a given flux, the jet is brighter in the OIR on the decline of the outburst compared to on the rise. It was shown \cite{russet07} that there is a hysteresis pattern between IR and X-ray emission from this source during the same outburst in that the declining hard state has a higher infrared flux at a given X-ray flux than the rising hard state. The OIR CMDs here indicate that there is indeed a different spectrum or radiative efficiency of the jet on the outburst decline, suggesting that this is the reason for the IR--X-ray hysteresis, and that it is unlikely to be due to changes in the X-ray evolution or the accretion viscosity parameter.

\begin{figure}
\includegraphics[width=15cm,angle=0]{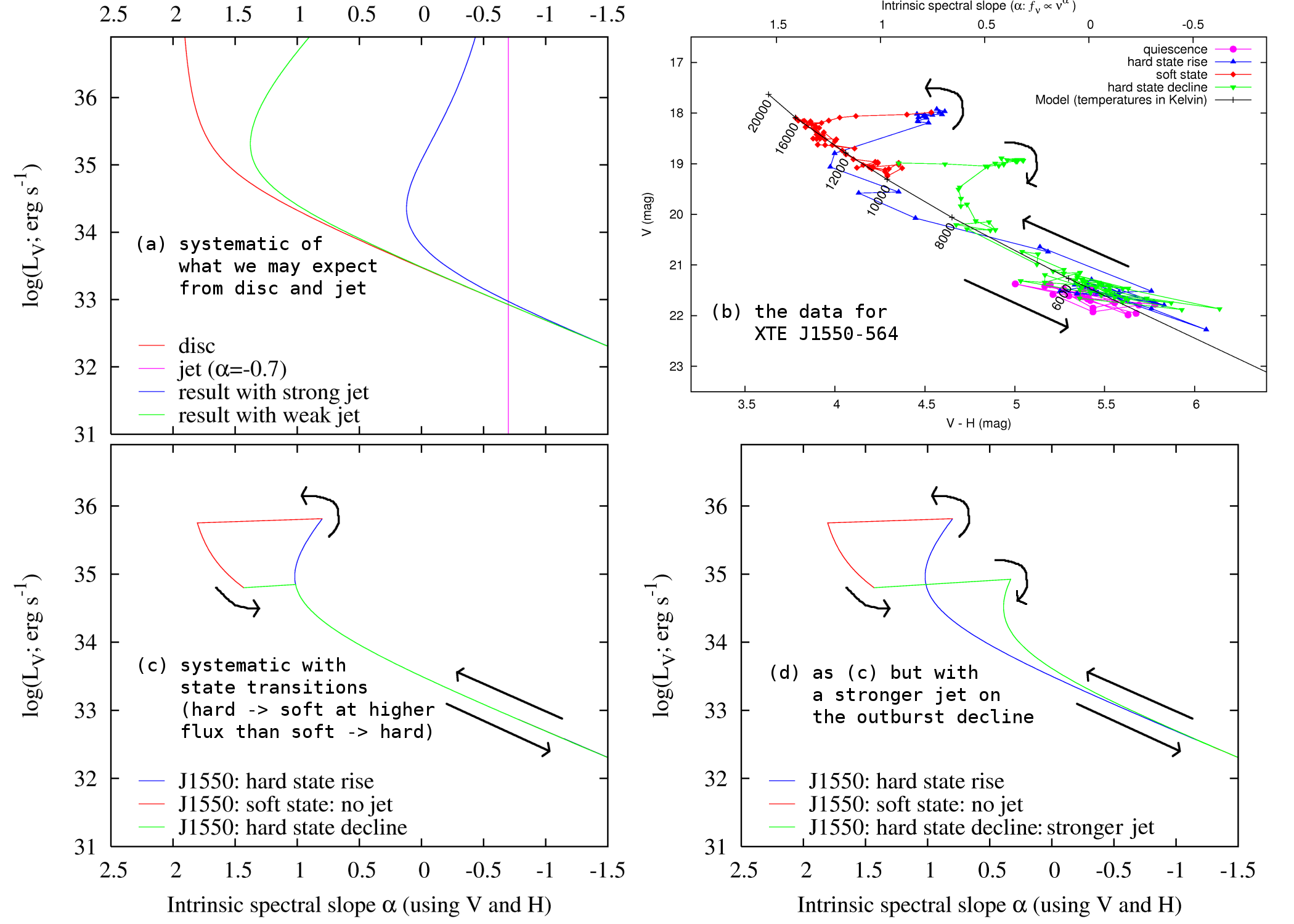}
\caption{Systematic CMDs assuming an OIR jet/disc ratio that increases with flux (see text). (a) The jet (no colour change) and irradiated disc (bluer at higher fluxes), and the combination of the two using two different normalisations for the jet. (b) The CMD of XTE J1550--564 (Fig. 1). (c) The systematic of the combination of jet and disc (as in (a)) with state transitions from/to the hard state to/from the soft state (the latter transition at a lower flux than the former). We assume the jet emission disappears in the soft state, creating a hysteresis in the CMD. (d) The same as (c) but with a higher normalisation for the jet on the decline compared to on the rise. The systematic in (d) approximately reproduces the empirical evolution of the outburst of XTE J1550--564 shown in (b).}
\end{figure}

\begin{figure}
\includegraphics[width=13cm,angle=0]{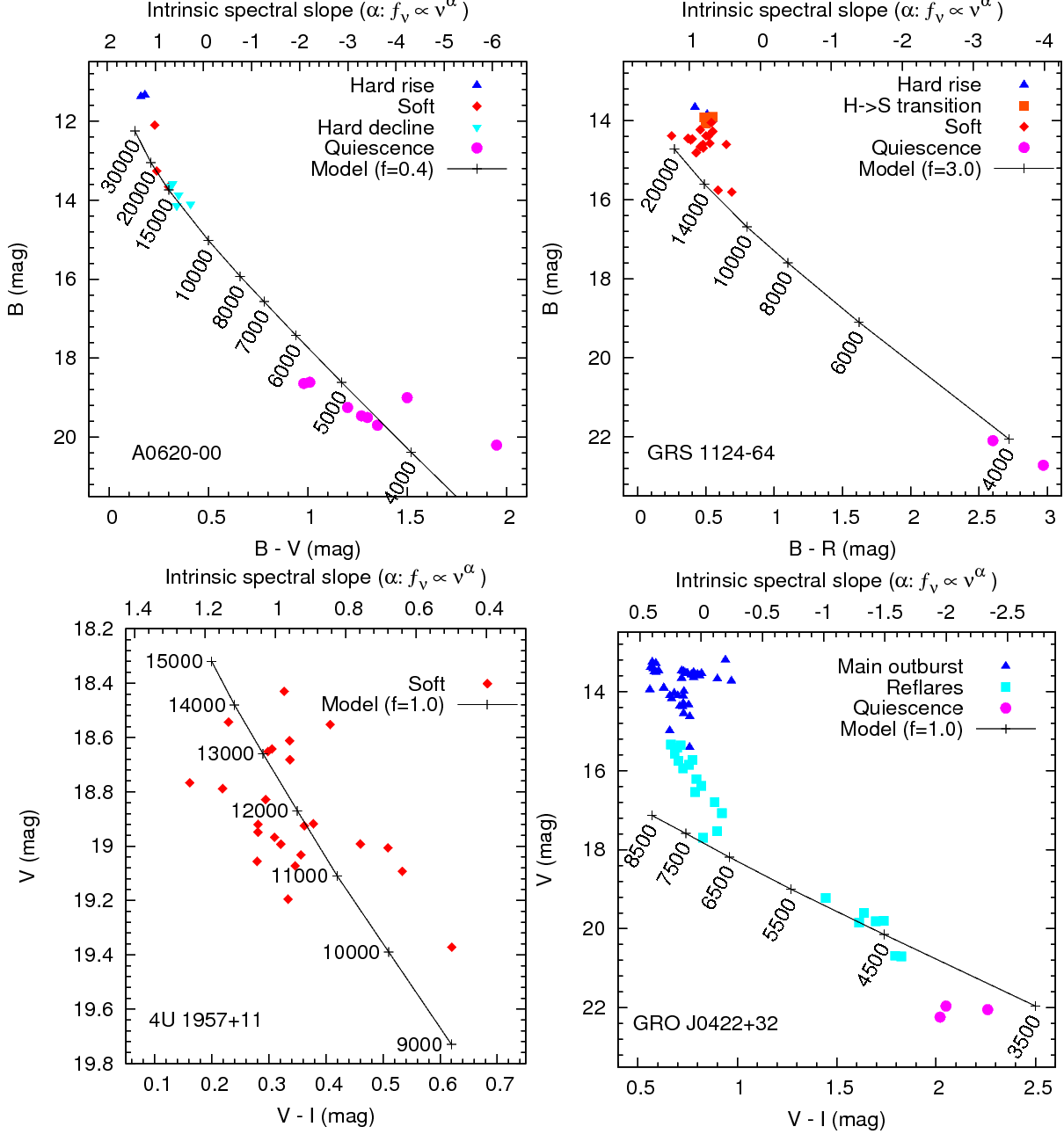}
\caption{CMDs of four black hole LMXBs (4U 1957+11 is most likely has a black hole primary but this is as yet unconfirmed). In general, blue data represent a hard X-ray state (upward arrows for outburst rises, downward for declines, squares for reflares), red data (diamonds) indicate a soft state, orange data (also squares) indicate transitions and purple data (circles) are data taken in quiescence. The magnitudes are taken from the literature except for 4U 1957+11, for which new observations have been made with the FT North (Russell et al. in preparation).}
\end{figure}

\begin{figure}
\includegraphics[width=13cm,angle=0]{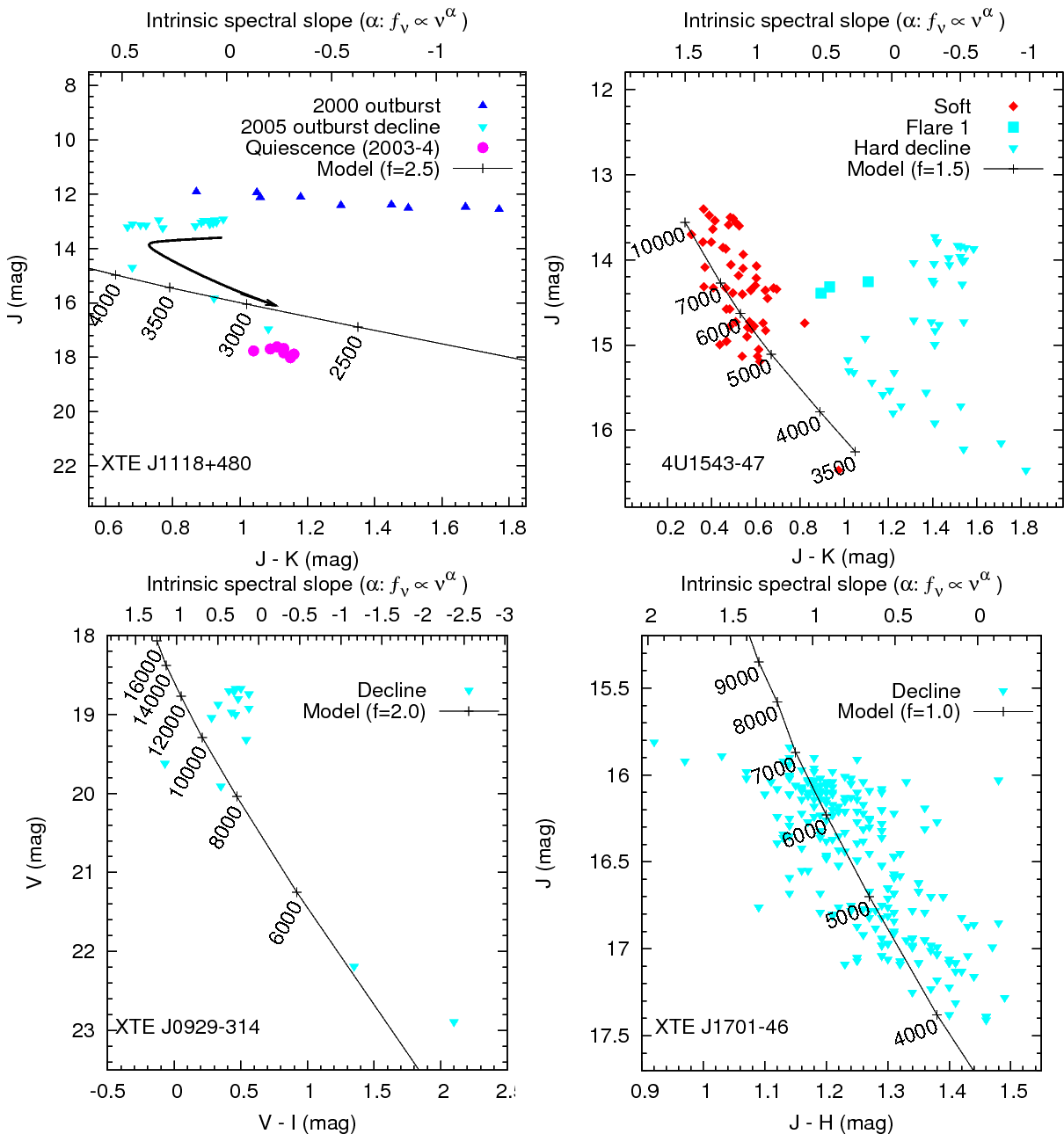}
\caption{CMDs of four LMXBs. The bottom two panels are the only neutron star XBs; all other CMDs are of black hole XBs. Colours and symbols are as in Fig. 3. The magnitudes are taken for the literature except for XTE J1701--462, for which new observations have been made with the Yale 1.0-m telescope \cite{maitba06} (see Russell et al. in preparation for details).}
\end{figure}

We present the CMDs of eight more LMXBs, with the blackbody model overlayed, in Figs. 3 and 4. The outbursts of A0620--00 and GRS 1124--64 can be approximated by the blackbody model over a range of eight magnitudes (optical wavebands were used for these two and not IR, so we expect the jet to contribute less in these CMDs). 4U 1957+11 is a persistent source and its magnitude range is just one magnitude, but the blackbody model is able to reproduce the slope of the relationship between colour and magnitude.

For GRO J0422+32, the data at low flux levels are well approximated by the blackbody model, but there is a huge excursion from the model of several magnitudes at high fluxes. Jet emission is one possibility, since this source remained in the hard state throughout its outburst, but the excursion away from the blackbody model is not as sharp as in the case of XTE J1550--564. Instead, the data reach an approximately constant spectral index of $0 < \alpha < +0.5$. This is consistent with the viscously heated disc, where we expect $\alpha \sim +0.3$, however according to the relations with the X-ray flux we would only expect the viscous disc to dominate at lower fluxes, not higher ones. An alternative explanation is the size of the emitting region of the blackbody changes. For example, near the peak of the outburst the outer disc may be warped or puffed up, or the outer radius of the disc may be larger. This could happen due to the intense X-ray radiation field or a very high mass accretion rate. The result would be a larger emitting area, but with the same temperature, meaning a higher flux but a similar colour. This has been documented in the past -- a study of the optical--UV spectral energy distributions (SEDs) of the black hole XB XTE J1859+226 \cite{hyneet02} implies an accretion disc whose area evolved during an outburst. OIR SEDs of GRO J0422+32 (BVRI-bands) indicate there is an I-band excess at high fluxes most likely from the jet, but it is not pronounced enough to explain the excursion of several magnitudes from the blackbody model.

The CMDs of two black hole XBs and two neutron star XBs are presented in Fig. 4. The black hole XB XTE J1118+480 has well known jets that emit in the OIR \cite{hyneet06}; here the data are much brighter/redder than the blackbody model, except at the lower fluxes near quiescence. This source has a short orbital period and so has a small accretion disc. Perhaps this implies a higher jet/disc ratio in this source compared to most long-period LMXBs, which have larger, brighter discs. For 4U 1543--47 the data during the hard state are redder than expected from the blackbody model whereas the soft state data lie close to the model. Interestingly for this source the data are still not consistent with the blackbody model at low fluxes in the hard state, but the companion star is bright in this source and the colour may be dominated by its influence at low flux levels.

The neutron star XB (and millisecond X-ray pulsar) XTE J0929--314 has a CMD which can be approximated by the model, but the data at the highest fluxes are too red. Indeed, this source, like most millisecond X-ray pulsars has a transient IR-excess above the spectrum of the disc, which is consistent with synchrotron emission from its jets \cite{gileet05}. This CMD is further evidence for jets dominating the OIR in millisecond X-ray pulsars (jets likely do not dominate the OIR in most neutron star XBs, but in millisecond X-ray pulsars the orbital periods are short and the accretion discs are small and faint, so the jet/disc ratio is likely higher for these systems). Finally, the blackbody model is able to reproduce the general relation between colour and magnitude in the neutron star XB XTE J1701--462.

\section{Multiwavelength disc--jet unification}

\begin{figure}
\centering
\includegraphics[width=9cm,angle=0]{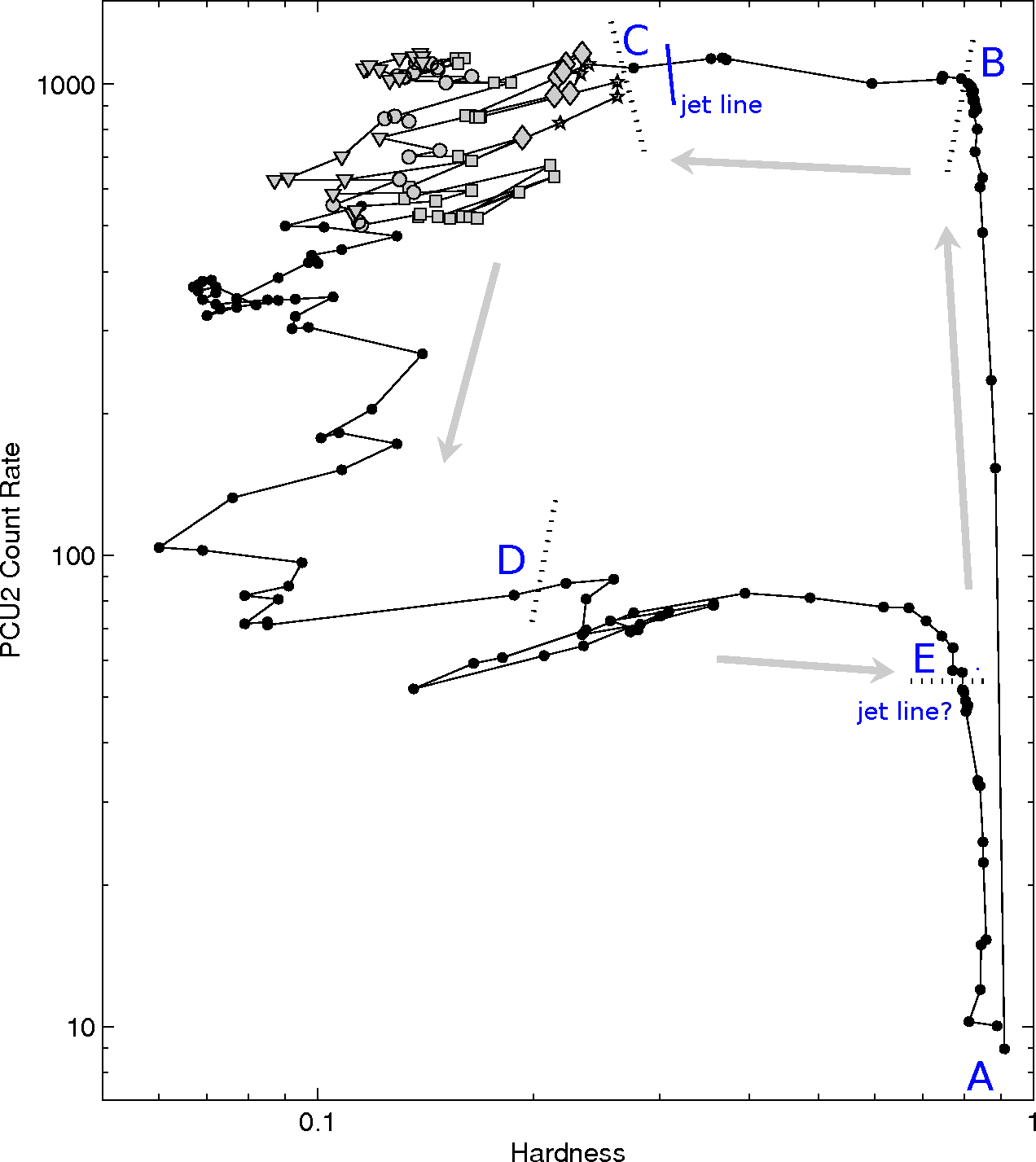}
\caption{X-ray HID of GX 339--4 (adapted from \cite{bellet05}).}
\end{figure}

A general picture of the X-ray behaviour of transient black hole XB outbursts has now emerged \cite{fendet04} in which a typical outburst follows a specific path in the X-ray HID (Fig. 5). Since the OIR fluxes and colours are correlated with the X-ray and radio luminosity, all these wavebands can now be included in the picture. This allows predictions of the behaviour of one waveband from the behaviour of another. At the beginning stages of a black hole XB outburst the source is in the hard state, and luminosity rises as $F_{\rm OIR} \propto F_{\rm radio} \propto F_{\rm X}^{0.6-0.7}$ (stage A to B in Fig. 5). During this time the jet/disc ratio in the OIR increases, and at some point the jet dominates over the accretion disc emission. Some sources do not make a state transition \cite{brocet04} and decline in luminosity back into quiescence. If a source instead makes a transition, the X-ray spectrum softens at B and the IR (and optical to some extent) emission starts to drop but the radio persists until the `jet line'. Here, a bright, optically thin radio flare is seen and the jet is thought to be then quenched at C. From C to D no radio emission from the steady jet (core) is detected and the X-ray spectrum is soft. The OIR colours are consistent with that of a cooling accretion disc, as the X-ray and OIR fluxes decrease. Radio emission is sometimes detected in the soft state from a fading optically thin source which can be spatially resolved (e.g. jet shock material); this is consistent with a component physically separated from the core. The source makes a transition back into the hard state at E and the steady jet returns either before or at E \cite{kaleet05} (Fender et al. in preparation). The IR from the jet appears to take a number of days to rise once back in the hard state \cite{jainet01,buxtba04}; this could be due to the spectrum of the jet becoming optically thick at higher frequencies with time, as the jet is formed and builds up (the jet may start off optically thin at radio frequencies). The OIR colour is then redder than expected from the disc, and in at least one source, the jet appears stronger (as implied by a higher IR flux) on the outburst decline compared to the rise. The radio, OIR and X-ray then all drop to quiescence following the correlations unless a secondary outburst is initiated in the mean time again by increased mass accretion. The OIR colours change progressively until at lower fluxes the disc dominates and the jet likely only dominates in the radio and mid-IR.

\section{Summary}

We have used OIR colour-magnitude diagrams of the outbursts of nine low-mass X-ray binaries to successfully separate thermal disc emission from non-thermal jet emission. A heated single-temperature blackbody is able to approximately reproduce the observed relation between colour and magnitude in all systems, except when excursions are made to a redder colour than expected. We find that this is due to a non-thermal jet component, which tends to dominate the optical and IR moreso in X-ray binaries with shorter orbital periods (i.e. the ones with smaller, fainter accretion discs). Most other data are consistent with an outer disc of constant area heated by illuminating X-ray photons. There is tentative evidence for the size of the accretion disc of GRO J0422+32 to change during its outburst.

The OIR colours and fluxes are correlated with X-ray and radio information, and so a universal behaviour is found. As a result, it is possible to some extent to predict the broadband emission properties of LMXBs from the monitoring of only one of the three: radio, OIR or X-ray (via the position in the X-ray HID). We find that X-ray state changes can be easily identified by OIR colour changes; the hard and soft X-ray states corresponding to redder and bluer OIR spectra (at the same flux level at least). There are fewer data for neutron star XBs, but similar X-ray hysteresis has been found \cite{maccco03} and the bright radio flares may also exist for these sources; none have been detected so far but neutron star-powered jets are generally dimmer and weaker than black hole-powered jets. Radio jets remain detectable in soft X-ray states of neutron star XBs \cite{miglfe06} so it would be interesting to test whether the IR jet emission drops in the soft state. IR observations would be required of a transient neutron star XB with a small accretion disc that performs state transitions. Millisecond X-ray pulsars are the perfect candidates except that they remain in a hard X-ray state.

\end{document}